\documentclass[pagesize,12pt,a4paper]{scrartcl}
\usepackage{german}
\usepackage{amsmath}
\usepackage{amsthm}
\usepackage{amssymb}
\usepackage{amsxtra}
\usepackage{graphicx}
\usepackage{braket}
\usepackage{capt-of}
\usepackage{ulem}
\usepackage{ae}
\usepackage{makeidx}
\usepackage{inputenc}
\usepackage[T1]{fontenc}
\usepackage{url}
\usepackage{eepic}
\usepackage{mathrsfs}
\usepackage{comment}
\usepackage{mparhack}
\newcommand{\ir}{\mathrm{i}}
\newcommand{\e}{\mathrm{e}}

\newcommand{\eins}{{\mathbf 1}}

\renewcommand*\emph[1]{{\textit{#1}}}
\renewcommand{\jmath}{j}
\newcommand{\longpage}{\enlargethispage{1\baselineskip}}
\newcommand*\idx[1]{\index{#1}#1}

\DeclareMathOperator{\ad}{ad}

\DeclareMathOperator{\dv}{d}

\DeclareMathOperator{\dt}{\tilde d\!}

\DeclareMathOperator*{\diag}{diag}
\ifx\KOMAScript\undefined%
  \DeclareRobustCommand{\KOMAScript}{\textsf{K\kern.05em O\kern.05em%
      M\kern.05em A\kern.1em-\kern.1em Script}}
\fi
\newlength{\help}
\newlength{\minuslaenge}
\newtheoremstyle{note}
  {3pt}
  {3pt}
  {\rmshape}
  {}
  {\bfseries}
  {:}
  {.5em}
  {}

\theoremstyle{note}

\makeatletter

 \def\vec#1{\ensuremath{\mathchoice
                     {\mbox{\boldmath$\displaystyle\mathbf{#1}$}}
                     {\mbox{\boldmath$\textstyle\mathbf{#1}$}}
                     {\mbox{\boldmath$\scriptstyle\mathbf{#1}$}}
                     {\mbox{\boldmath$\scriptscriptstyle\mathbf{#1}$}}}}%
\makeatother

\begin{document}

  \title{The Rough with the Smooth of the Light Cone String}
  \author{Norbert Dragon and Florian Oppermann\\
          Institut f\"ur Theoretische Physik\\
          Leibniz Universit\"at Hannover 
}
\date{}

\maketitle

\begin{abstract} 
The polynomials in the generators of a unitary representation of the Poin\-car\'e group constitute an algebra 
which maps the dense subspace $\mathcal D$ of smooth, rapidly decreasing wavefunctions to itself.
This mathematical result is highly welcome to physicists,
who previously just assumed their algebraic treatment of unbounded operators be justified.
The smoothness, however, has the side effect that a rough operator $R$, which 
does not map a dense subspace of $\mathcal D$ to itself, has to be shown to allow for some other
dense domain which is mapped to itself both by $R$ and all generators.
Otherwise their algebraic product, their concatenation, 
is not defined.

Canonical quantization of the light cone string postulates operators $-\ir X^1$ and
$P^-=(P^0 - P^z)/2$ and as their commutator the multiplicative operator $R=P^1/(P^0 + P^z)$. 
This is not smooth but rough on the negative $z-$axis of massless momentum.

Using only the commutation relations of $P^m$ with the generators $-\ir M_{iz}$
of rotations in the $P^i$-$P^{z}$-plane we show that on massless states the operator~$R$
 is inconsistent with a unitary representation of SO$(D-1)$.
This makes the algebraic determination of the critical dimension, $D=26$, of the
bosonic string meaningless: if the massless states of the light cone string admit $R$ then they 
do not admit a unitary representation of the subgroup SO$(D-1)$ 
of the Poincar\'e group.

With analogous arguments we show: Massless multiplets are inconsistent with a 
translation group of the spatial momentum which is generated by a self-adjoint spatial position operator 
$\vec X$.

\end{abstract}

\newpage

\section{Introduction}

The seminal calculation \cite{goddard} of the critical dimension $D=26$ of the spacetime,
 in which the bosonic quantum light cone string acts, left two nagging doubts.

If, assuming some basic algebraic rules, formally hermitian operators $M_{mn}=-M_{nm}$, $m,n \in \set{0,1,\dots D-1}$, 
satisfy the commutation relations of the Lorentz Lie algebra,\footnote{In an orthonormal basis our 
metric  is $\eta = \diag (1,-1,\dots, -1)$.} 
\begin{equation}
\label{loralgebra}
 [-\ir M_{mn}, -\ir M_{rs}] = -\eta_{mr}(-\ir M_{ns})+\eta_{ms}(-\ir M_{nr})+\eta_{nr}(-\ir M_{ms})-\eta_{ns}(-\ir M_{mr})
\end{equation}
is this also sufficient for the operators to generate a unitary representation of the Lorentz group? Each finite dimensional matrix $\omega$  
generates by the exponential map the elements 
of a one parameter group, 
\begin{equation}
g_t=\e^{t\,\omega} \ ,\ t\in \mathbb R\ ,\ g_t g_{t'}=g_{t+t'}\ .
\end{equation}
By the Baker Campbell Hausdorff formula exponentials of sufficiently small finite dimensional matrices~$\omega$  
generate a  corresponding Lie group $G$, if the matrices $\omega$ represent a Lie algebra~$\mathfrak g$.
But which requirements are there for unbounded operators? Is a Lorentz Lie algebra in $D=26$ of formally skew-adjoint 
generators sufficient for the existence of a unitary representation of the Poincar\'e group?

Vice versa, if a set of operators does \emph{not} satisfy the commutation relations of the Lorentz Lie algebra does this exclude improved operators which do so? 
In this case $D=26$ would be indicated only as the dimension where an apparent but correctable anomaly vanishes. 
Is $D=26$ necessary for the Lorentz invariance of the light cone string?

Neither of these questions could be addressed seriously. The cumbersome calculation of the Lorentz Lie algebra shied away all attempts
to exponentiate the generators or to investigate classes of improvement terms. By common consensus, one neglected the existence problem 
of exponentiated generators and adopted $D=10$, the critical dimension
of the superstring, as start value for compactification schemes.

We shortly review the main results of the mathematical investigations \cite{schmuedgen} of operators
which generate in Hilbert space a unitary representation of a finite dimensional Lie group.
They allow us to deduce the results announced in the abstract. Moreover, we point out that
the excitation operators of the light cone string $\alpha_{-l}$, $l\in \mathbb N$, which map the tachyon shell and the massless shell 
in a momentum local way to massive shells, cannot map smooth wavefunctions of the tachyon shell and the 
massless shell to smooth wavefunctions of massive shells.

Though similar and even more severe inconsistencies exist with tachyon states, 
we restrict our considerations mainly to problems with massless states. They persist 
even if one could get rid of the tachyon.

\section{Smoothness of Lie Group Transformations}

Let $U_g:\mathcal H\rightarrow \mathcal H $ denote a unitary representation $U_g U_{g'} = U_{g\, g'}$ of a Lie group $G$ 
in a Hilbert space~$\mathcal H$ for which
all maps $f_{[\Psi]}: g \mapsto U_g \Psi $ from $G$ to $\mathcal H$ are measurable. 

The generators $-\ir M_\omega$ of one-parameter subgroups $U_{\e^{t\omega}}$  
are defined on the subspace of smooth states $\Psi\in \mathcal D\subset \mathcal H$ on which all $U_{\e^{t\omega}}$ act differentiably, 
\begin{equation}
-\ir M_\omega \Psi = \lim_{t\rightarrow 0} \frac{U_{\e^{t\omega}}\Psi - \Psi}{t}\ .
\end{equation}

By Stone's theorem each $M_\omega$ is essentially self-adjoint. It owns a~\idx{projection valued measure}\footnote{A 
projection valued measure $E_\lambda$ is a parameterized set of projectors, $\lambda \in \mathbb R$, 
with $E_\lambda E_{\lambda'}= E_{\min\set{\lambda.\lambda'}}$, 
$\lim_{\varepsilon \rightarrow 0+}E_{\lambda+\varepsilon}= E_\lambda $, $E_{-\infty}=0$ and $E_\infty=\eins$.}
by which it generates $U_{\e^{t\omega}}$ not only in $\mathcal D$ but in the complete Hilbert space $\mathcal H$, 
\begin{equation}
M_\omega = \int\! \dv\! E_{\lambda}\,\lambda\ ,\ 
U_{\e^{t\,\omega}} = \int\! \dv\! E_{\lambda}\,  \e^{-\ir\,  t\,\lambda}=:\e^{-\ir t M_\omega}\ .
\end{equation}

Applied to states in $\mathcal D$ the products 
\begin{equation}
\label{ug}
U_{g(t)}=U_{\e^{t_1\omega_1}}\cdots U_{\e^{t_n\omega_n}}\ ,\ t=(t_1,t_2,\dots t_n)\in \mathbb R^n
\end{equation}
are a differentiable function of $\mathbb R^n$. So the derivatives 
\begin{equation}
\partial_{t_1}\dots \partial_{t_n}U_{g(t_1\dots t_n)_{|_{t=0}}} = (-\ir)^n M_{\omega_1}\cdots M_{\omega_n}
\end{equation}
exist on them no matter how large~$n$ is. The concatenation of generators, their algebraic product, is defined 
in $\mathcal D$. It is a domain of the polynomial algebra $\mathcal A$ of the generators~$M_\omega$ 
(which restricted to $\mathcal D$ are essentially self-adjoint)
and invariant also under all $U_g,\,g\in G$. The generators represent the corresponding Lie algebra  $\mathfrak g$ \cite{schmuedgen}: 
\begin{equation}
\label{replie}
M_\omega M_{\omega'}-  M_{\omega'} M_\omega - \ir  M_{[\omega,\omega']}
\end{equation}
is a two-sided ideal which vanishes in the algebra $\mathcal A$ if multiplied from the left and the right 
with  arbitrary  polynomials $A_l, A_r \in \mathcal A$.

As the maps $f_{[\Psi]}:g \mapsto U_g \Psi $ are measurable for all $\Psi$
and because integrals over measurable functions of \emph{compact} support exist,
therefore the G\aa rding space $\mathcal G$ exists which is 
spanned by smoothened states $\Psi_f$ which are averaged with a left invariant
volume form~$\dv\! \mu_g = \dv\! \mu_{g'g}$ and smooth functions $f: G \rightarrow \mathbb C$ of compact support
\begin{equation}
\label{smoothing}
\Psi_f = \int_{G}\! \dv\! \mu_g \, f(g)\, U_g \Psi\ .
\end{equation}
The smoothened states $\Psi_f$ transform smoothly
\begin{equation}
\label{smoothtrans}
U_g \Psi_f = \Psi_{f\circ g^{-1}}\ .
\end{equation}

The G\aa rding space coincides with the space of smooth states,  $\mathcal G = \mathcal D$ \cite{schmuedgen}. 
It is dense in the Hilbert space $\mathcal H(G)$ of square integrable functions of $G$. 
That $\mathcal G$ exists and constitutes the dense and invariant domain of the polynomial algebra of the generators and the group,
justifies in retrospect physicists who manipulated the unbounded generators $M_\omega$ algebraically not caring about domains. 

A mass shell is the Lorentz orbit  $G/H$ of some chosen momentum $\underline p$ with $p^0 > 0$
 and $\underline p^2 = m^2$.
Its smooth, rapidly decreasing wavefunctions also span a dense subspace $\mathcal D\subset \mathcal H(G/H)$
which is mapped to itself  by the unitary representation of the 
Poincar\'e group. Because its states are smooth, differential geometry becomes applicable to quantum physics.

Recall that a Hilbert space of square integrable functions $\Psi: p \mapsto \Psi(p)$ consists more precisely of equivalence classes of functions,
where functions are equivalent, if the support of their difference has measure zero.
\begin{equation}
\Psi = 0 \Leftrightarrow \braket{\Psi | \Psi}=\int\!\!\dv \!p\, |\Psi(p)|^2 = 0
\end{equation}
So the values in a set of measure zero do not count for equivalent functions.
A smooth function, however, is the only smooth function in its equivalence class. It is determined and smooth \emph{everywhere} not only \lq almost every\-where\rq .

The rough which one has to take with the smooth:
If in a relativistic model the commutation relations, which are postulated in canonical quantization, 
contain a rough ope\-rator $R$, which does not map all smooth states to smooth states, 
then one has to check, whether this is consistent with the unitary representation $U_{a,\Lambda}$ of
the Poincar\'e group.\footnote{Notation: 
Let $T_{a}: x \mapsto x + a$ denote a translation in $\mathbb R^4$, $T_\Lambda: x \mapsto \Lambda x$ a Lorentz transformation, 
and $T_{a,\Lambda}=T_a T_\Lambda \in \mathfrak P$ a Poincar\'e transformation.
We denote by $U_{a,\Lambda}$ its unitary representation with generators $P^m$ and $M_{mn}$, 
$U_{a,\e^\omega}= \e^{\ir P a}\e^{\ir \omega^{mn }M_{mn}/2}$,
in a Hilbert space $\mathcal H_1$ of one-particle states. }  

For this consistency we require $R$ and all its commutators 
with generators of the Poincar\'e group ($U_{g(t)}$ given by (\ref{ug}))
%
\begin{equation}
\label{adrough}
\ad_{M_{\omega_1}} \dots \ad_{M_{\omega_n}} R 
:= \ir^n \bigl(\partial_{t^1}\dots \partial_{t^n}\bigr)_{|_{t=0}} U_{g(t)} R\, U_{g(t)}^{-1}
=[M_{\omega_1}, \dots [M_{\omega_n},R]\dots ]]
\end{equation}
to exist on all states $\Psi\in \mathcal D$.

Earlier arguments concerning a rough operator used the assumption that $R$ and its commutators 
with generators $M_\omega$ were elements of a common operator \emph{algebra,} a condition which underlies
implicitly all algebraic calculations and which is postulated explicitly at the very beginning of the
mathematical investigation \cite{schmuedgen}. There the definition of an operator algebra postulates 
the existence of a dense domain on which all operators are defined and which is left invariant by all
operators.

These arguments we criticized objecting that requiring an operator algebra may be a too strong 
assumption. Only the operator $R$ and the algebra of the generators $M_\omega$ on its right and on 
its left have to exist. There could be a set of rules, though presently unknown, to justify the 
calculation of e.g. $D=26$ but rule out the inconsistency 
which results from assuming an operator algebra of $R$ and the generators.

Because of these objections we show the inconsistency of the example of 
the multi\-plicative operator $R=P^1/(P^0 + P^z)$
based solely on the assumed existence of 
all repeated commutators of $R$ with generators of rotations, if
applied to massless, smooth states. 

\section{Momentum Local Maps}

Specific to the gauge fixed quantum string are excitation operators $\alpha_{-l}$, $l\in \mathbb N$, which
excite states on mass shells $m^2(N)= (N-1)\,\mu^2$ to states on mass
shells $m^2(N+ l)$ \cite{arutyunov,thooft}.
The transition is momentum local such that for each momentum with $p^2=m^2(N)$ there is an excited momentum $q=g_l(p)$ with  $q^2= m^2(N+l)$ and
\begin{equation}
\label{phil}
\Phi_l(q) := (\alpha_{-l}\Psi)(q) = \sqrt{\frac{q^0}{p^0}\bigl|\det\!
 \frac{\partial g_l^{-1}}{\partial q}\bigl|\,}  M_l(p) \Psi(p)\ ,\ p=g_l^{-1}(q)\ .
\end{equation}
$M_l(p)$ is some matrix, unitary on its image, with indices which we need not depict. 
We have made the factor explicit, which relates the measures $\dt p\sim \dv^{D-1}/p^0$ and $\dt q$,
because it vanishes where $g_l$ has no inverse and is a rough multiplicative operator where $g_l^{-1}$ does not
exist. By their commutation relations and the ground state property $\alpha_l \Psi = 0$, $\alpha_l = (\alpha_{-l})^\star$, $l\ge 2$, for massless states~$\Psi$,
the operators $\alpha_{-l}$ can be inverted on their image,
\begin{equation}
\label{psip}
\Psi(p) = (\alpha_{l}\Phi_l)(p) = \sqrt{\frac{p^0}{q^0}
\bigl|\det\! \frac{\partial g_l}{\partial p}\bigl|\,}
M_l(p)^{-1} \Phi_l(q)\quad \text{(no sum over }l),\ q=g_l(p)\ .
\end{equation}

These operators are potentially inconsistent with the Poincar\'e generators. They act on smooth functions 
of the massless shell $\mathcal M_0=\set{p: p= \e^{\lambda}\,(1,\vec n),\, \vec n \in S^{D-2},\, \lambda \in \mathbb R}$ 
or the tachyon shell $\mathcal M_{\text{Tachyon}}=\set{p: p= (E, \sqrt{\mu^2 + E^2}\,\vec n),\,  \vec n \in S^{D-2},\, E \in \mathbb R}$
which have the topology of  $ S^{D-2}\times \mathbb R$.
Each massive shell $\mathcal M_m= \set{p: p^0=\sqrt{m^2+\vec p^2},\vec p \in \mathbb R^{D-1}}$ has the topology of  $\mathbb R^{D-1}$. 

For $\alpha_{-l}$, $l\ge 2$, to map the smooth tachyonic and massless wave functions to smooth massive wave functions, the momentum map
$g_l$ has to map the tachyon and the massless shell smoothly and with a smooth inverse  $g_l^{-1}$ to massive shells.
But the topologies of the shells are different:
there is no diffeomorphism $g_l$ of $ S^{D-2}\times\mathbb R $ to $\mathbb R^{D-1}$. 

\begin{figure}[ht]
\centering
\includegraphics[scale=1.2]{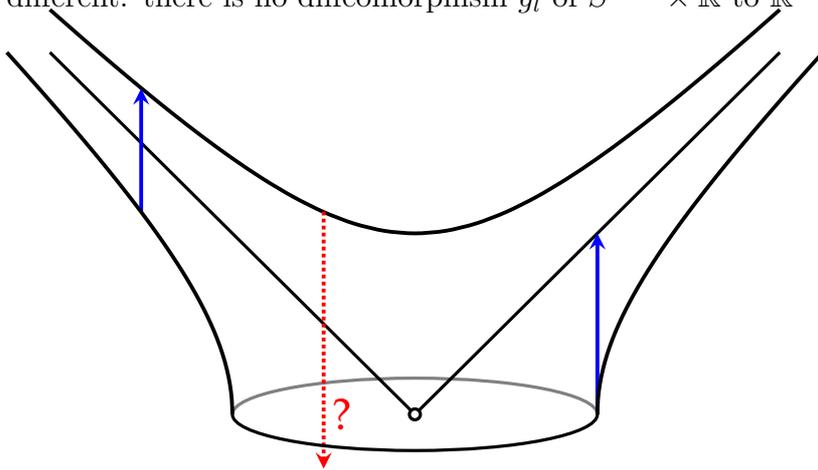} 
\caption{Transitions between Mass Shells in the Static Gauge}
\label{staticgauge}
\end{figure}

Hence, in a relativistic theory it is highly questionable whether operators $\alpha_{-l}$ exist
which excite a tachyon or massless particles in a momentum local way to massive particles.

In fig.\ref{staticgauge} the failure of invertible transitions between the massive, massless and tachyon shell is obvious. 
In the  static gauge of the  bosonic string \cite{jorjadze}
they transfer only energy. But massive states with spatial momentum $|\vec p| < \mu $
are not related to tachyon states.


No one can do better: the different topologies exclude  any smooth,
invertible map between a massive shell and the massless or the tachyon shell.


\section{Lie Algebra without and with Group}
The mere fact, that differential operators satisfy a Lie algebra on some space of functions does not make them generators of a representation of the
corresponding group. This is demonstrated by the following operators $M_{mn}=-M_{nm}$,
\begin{equation}
\label{masselosnord}
\begin{aligned}
\bigl(-\ir M_{12}\Psi\bigr)_N(p)&= - \bigl(p_x\partial_{p_y} - p_y\partial_{p_x}\bigr)\Psi_N(p) - \ir\, h\,\Psi_N(p)\ ,\\
\bigl(-\ir M_{31}\Psi\bigr)_N(p)&= - \bigl(p_z\partial_{p_x} - p_x\partial_{p_z}\bigr)\Psi_N(p) - \ir\, h\,\frac{p_y}{|\vec p| + p_z}\Psi_N(p)\ ,\\
\bigl(-\ir M_{32}\Psi\bigr)_N(p)&= - \bigl(p_z\partial_{p_y} - p_y\partial_{p_z}\bigr)\Psi_N(p) + \ir\, h\,\frac{p_x}{|\vec p| + p_z}\Psi_N(p)\ ,\\
\bigl(-\ir M_{01}\Psi\bigr)_N(p)&=  |\vec p| \partial_{p_x} \Psi_N(p) -\ir\,h\,\frac{p_y}{|\vec p| + p_z}\Psi_N(p)\ ,\\
\bigl(-\ir M_{02}\Psi\bigr)_N(p)&=  |\vec p| \partial_{p_y} \Psi_N(p) +\ir\,h\,\frac{p_x}{|\vec p| + p_z}\Psi_N(p)\ ,\\
\bigl(-\ir M_{03}\Psi\bigr)_N(p)&=  |\vec p| \partial_{p_z} \Psi_N(p)\ .
\end{aligned}
\end{equation}
On differentiable functions of the northern coordinate patch $\mathcal U_N$ of the massless shell $\mathcal M_0$
\begin{equation}
\mathcal U_N = \set{p: p^0=\sqrt{\vec p^2}\,,\, |\vec p| + p_z > 0}\subset \mathcal M_0=\set{p: p^0 = \sqrt{\vec p^2} > 0 }\subset \mathbb R^4
\end{equation}
the operators $-\ir M_{mn}$ satisfy the Lorentz Lie algebra\footnote{Observe
 $\sum_{i=1}^{2}p^ip^i=|\vec p |^2-(p_z)^2=(|\vec p |+p_z)(|\vec p|- p_z)$.} (\ref{loralgebra}) 
in $D=4$ \cite{bose, ek, fronsdal, lomont}.
The angular momentum in the direction of the momentum, the helicity~$h$, 
\begin{equation}
\label{helizit}
\bigl(({p_x}\,M_{23}+ {p_y}\,M_{31}+ {p_z} \,M_{12})\Psi\bigr)_N(p) = h\,|\vec p|\,\Psi_N(p)
\end{equation}
is some real number. The Lorentz Lie algebra does not restrict $2h$ to be an integer.

The operators are formally skew-adjoint with respect to the Lorentz invariant measure $\dt p = \dv^{3}\!p / |\vec p|$,  formally only, because
the singularities at $|\vec p|+ p_z = 0$ need closer investigation.

The operators (\ref{masselosnord}) cannot generate a unitary representation of the Lorentz group 
because the domain $\mathcal U_N$ of the differentiable functions is too small: Lorentz generators  act on 
smooth states, which have to be defined \emph{everywhere} in the Lorentz orbit $\mathcal M_0$.
The group acts transitively on the massless shell  and contains e.g. for each massless momentum $p$ a rotation 
which maps $\vec p$ to the negative $z$-axis
\begin{equation}
\mathcal A_- = \set{p:p^0 = \sqrt{\vec p^2}\,,\,p^0 + p_z = 0}\ . 
\end{equation}

For negative $p_z$ and with $x = (\sum_{i=1}^{D-2}p_i^2) /p_z^2$, $p_z :=p^{D-1}$, one has for each $D$ 
\begin{equation}
\text{for }p_z < 0\,:\ 
|\vec p| + p_z=|\vec p| - |p_z|= |p_z|(\sqrt{1+x}-1)\le |p_z|\, \frac{x}{2}\ ,
\end{equation}
because the concave function $x\mapsto \sqrt{1+x}$ is bounded by its tangent at $x=0$. 
So
\begin{equation}
\label{lowerbound}
\text{for }p_z < 0\,:\ 
\frac{1}{|\vec p| + p_z}  \ge 
 \frac{2|p_z|} {r^2}\ ,\   r^2=\sum_{i=1}^{D-2}p_i^2 > 0\ ,
\end{equation}
diverges in a neighbourhood $\mathcal U$ of $\hat p \in \mathcal A_-$ at least like the inverse square of the 
axial distance to $\mathcal A_-$.  

If $h\,\Psi_N(\hat p)\ne 0$ then it must not be differentiable there. 
Otherwise the multiplicative term of  $M_{31}\Psi_N$ dominates near $\hat p$ where it scales as $ |p_z| / r$. 
Its squared modulus integrated  with $\dv^3\!p / |\vec p|$ over a sufficiently small $\mathcal U$ in cylindrical coordinates is bounded from below by a positive number
times an $r$-integral $r / r^2 \dv\! r$ which diverges at the lower limit $r=0$. The multiplicative term alone diverges.

Near $\mathcal A_-$ the derivative term $D\Psi_N= -p_z \partial_{p_x}\Psi_N$ in $M_{31}\Psi_N$ has to cancel 
the multiplicative singularity $M \Psi_N$ up to a function~$\chi$, which is smooth. 
This linear inhomogeneous condition $(D + M)\Psi_N=\chi$ is solved by variation of constants $\Psi_N = f \Psi_S$ where $f$ 
satisfies the two homogeneous conditions
\begin{equation}
|p_z|\bigl(\partial_{p_x} - 2 \ir h \frac{p_y}{p_x^2+p_y^2}\bigr)f= 0\ ,\ |p_z|\bigl(\partial_{p_y} + 2 \ir h \frac{p_x}{p_x^2+p_y^2}\bigr)f= 0\ ,
\end{equation}
for both $M_{31}\Psi_N$ and  $M_{32}\Psi_N$ to exist. 
They determine $f(p)=\e^{-2\ir h \varphi(p) }$ up to a factor. 

The function $\Psi_S$ is smooth in the southern coordinate patch
\begin{equation}
\mathcal U_S = \set{p: p^0=\sqrt{\vec p^2}\,,\, |\vec p| - p_z > 0}
\end{equation}
and related in $\mathcal U_N\cap \mathcal U_S$ by the transition function $f^{-1}=h_{SN}$ to $\Psi_N$
\begin{equation}
\label{suednordwell}
\Psi_S(p) = h_{SN}(p)\,\Psi_N(p)\ ,\ h_{SN}(p) = \e^{2\ir\, h\varphi(p)}=\Bigl (\frac{p_x+\ir p_y}{\sqrt{p_x^2+p_y^2}}\Bigr)^{2h}\ .
\end{equation}
The transition function $\e^{2\,\ir\,h\,\varphi(p)}$ is defined and smooth in $\mathcal U_N\cap \mathcal U_S$
only if $2h$ is integer. This is why the helicity of a massless particle  is integer or half integer.


Multiplying (\ref{masselosnord}) with $h_{SN}$ one obtains from  (\ref{suednordwell}) 
\begin{equation}
\label{masselossued}
\begin{aligned}
\bigl(-\ir M_{12}\Psi\bigr)_S(p)&= - \bigl(p_x\partial_{p_y} - p_y\partial_{p_x}\bigr)\Psi_S(p) + \ir\, h\,\Psi_S(p)\ ,\\
\bigl(-\ir M_{31}\Psi\bigr)_S(p)&= - \bigl(p_z\partial_{p_x} - p_x\partial_{p_z}\bigr)\Psi_S(p) - \ir\, h\,\frac{p_y}{|\vec p| - p_z}\Psi_S(p)\ ,\\
\bigl(-\ir M_{32}\Psi\bigr)_S(p)&= - \bigl(p_z\partial_{p_y} - p_y\partial_{p_z}\bigr)\Psi_S(p) + \ir\, h\,\frac{p_x}{|\vec p| - p_z}\Psi_S(p)\ ,\\
\bigl(-\ir M_{01}\Psi\bigr)_S(p)&=  |\vec p| \partial_{p_x} \Psi_S(p) +\ir\,h\,\frac{p_y}{|\vec p| - p_z}\Psi_S(p)\ ,\\
\bigl(-\ir M_{02}\Psi\bigr)_S(p)&=  |\vec p| \partial_{p_y} \Psi_S(p) -\ir\,h\,\frac{p_x}{|\vec p| - p_z}\Psi_S(p)\ ,\\
\bigl(-\ir M_{03}\Psi\bigr)_S(p)&=  |\vec p| \partial_{p_z} \Psi_S(p)\ .
\end{aligned}
\end{equation}

$\Psi_N$ and $\Psi_S$ are local sections of a bundle over $S^2\times \mathbb R$ with transition function $h_{SN}$. A massless quantum state $\Psi$
is a section given locally in $\mathcal U_N$ by $\Psi_N$ and in $\mathcal U_S$ by $\Psi_S$ \cite{dragon}.


All $M_{mn}\Psi$ are square integrable, rapidly decreasing and smooth in $\mathcal M_0$
if $\Psi$ is. 


For all $\omega$ in the Lorentz algebra  the operators $-\ir M_\omega = - \ir/2\, \omega^{mn}M_{mn}$ are by construction \cite{dragon}
the derivatives of unitary one-parameter  groups 
\begin{equation}
\label{sufficient}
-\ir M_\omega \bigl(U_{\e^{t\omega}} \Psi\bigr) = \partial_t \bigl(U_{\e^{t\omega}} \Psi\bigr)\ ,
\end{equation}
which act on a dense and invariant domain $\mathcal D$ of smooth states, where the transformations $U_{\e^{\omega}}$ together with all their products
represent unitarily the Lorentz group. So $-\ir M_\omega$ not only satisfy the Lorentz algebra but they are also
skew-adjoint (by Stone's theorem) and generate a unitary representation of the Lorentz group.

The singularities of the multiplicative terms in (\ref{masselosnord})
for $h\ne 0$ on the $3$-axis do not allow to conclude that the operators are rough.
They can be combined with the partial derivatives to \idx{covariant derivative}s 
\begin{equation}
\label{covdiv}
D_i  = \ir\, |\vec p|^{-1/2}\, M_{0i}|\vec p|^{-1/2} = \partial_{p^i}+ A_i -\frac{p^i}{2|\vec p|^{2}} \ ,
\end{equation}
where the connection $\vec A$ 
is given in the northern and southern coordinate patch by 
\begin{equation}
\label{ablcov}
\vec A_N(p)=
 \frac{\ir\,h}{|\vec p|(|\vec p|+p_z)}
\begin{pmatrix}
-p_y\\
\phantom{-}p_x\\
\phantom{-}0
\end{pmatrix} 
\ ,\ 
\vec A_S(p)=
\frac{-\ir \,h}{|\vec p|(|\vec p|-p_z)}
\begin{pmatrix}
-p_y\\
\phantom{-}p_x\\
\phantom{-}0
\end{pmatrix}\ .
\end{equation}
In the intersection of their coordinate patches they are related  by the transition function
\begin{equation}
D_{S\,i}=  \e^{2\,\ir\,h\,\varphi(p)} D_{N\,i} \,\e^{-2\,\ir\,h\,\varphi(p)}\ .
\end{equation}

The covariant derivatives $D_j$ and the momenta $P^i$  do not constitute Heisenberg pairs\index{Heisenberg pair} as 
the commutators $[D_i,D_j]$ do not vanish but yield the field strength of a momentum space monopole of charge $h$ at $p=0$,
\begin{equation}
\label{monopol}
\ [P^i,P^j] = 0\ ,\ [P^i,D_j] = -\delta^i{}_j\ ,\ 
\ [D_i, D_j]= F_{ij} = \partial_i A_j - \partial_j A_i 
= \ir\, h\,\varepsilon_{ijk}\frac{P^k}{|\vec P|^3}\ .
\end{equation}
The geometry of the massless momentum shell of particles with nonvanishing helicity is noncommutative. 

In terms of the covariant derivative the generators of Lorentz transformations 
(\ref{masselosnord}, \ref{masselossued}) take the smooth, rotation equivariant form ($\vec p \ne 0$)
\begin{equation}
\label{masseloskovariant}
-\ir M_{ij} = -(P^i D_{j}- P^j D_{i}) - \ir\,h\,\varepsilon_{ijk}\frac{P^k}{|\vec P|}\ ,\ 
-\ir M_{0i} = - |\vec P|^{1/2} D_{i}|\vec P|^{1/2}\ .
\end{equation}
They satisfy the Lorentz algebra on account of (\ref{monopol}) for arbitrary real~$h$.
However, the covariant derivative~$D$ is a skew-adjoint operator only if $2h$ is integer.

The integrand $\mathbb F = \frac{1}{2}\dv p^i\dv p^j F_{ij}$, the first Chern class, is a topological
density: integrated on each surface $\mathcal S$  which encloses the apex $p=0$ of the cone $p^0=|\vec p|$
\begin{equation}
\frac{1}{4\pi}\int_{\mathcal S}\mathbb F =  \ir\,h
\end{equation}
it yields a value which depends only on the transition functions of the bundle. The integral remains constant under smooth changes
of the connection $A$ of the covariant derivative as the Euler derivative of $\mathbb F(A)$ with respect to~$A$ vanishes. 

\section{Failing Rotational Symmetry of the Light Cone String}
\label{failing}

Canonical quantization of the light cone string \cite[page 23]{arutyunov,thooft} postulates 
transverse Heisenberg pairs $P^i$, $X^j$, $i,j \in \set{1,\dots D-2}$, which commute with  
$P^+ = (P^0 + P^z)$ and the level operator $N$,
\begin{equation}
\label{heisentrans}
 [P^i, X^j]= -\ir \delta^{ij}\ ,\ [P^i, P^j]= 0\ ,\  [X^i, X^j] = 0 \ ,\ [X^i, P^+] = 0\ ,\ [X^i,N]=0\ .
\end{equation}
By the mass shell relation 
\begin{equation}
P^- = \frac{1}{{2}}(P^0 - P^z) = \frac{(N-1)\mu^2 + \sum_{i=1}^{D-2} P^iP^i}{2 P^+}
\end{equation}
and the innocent looking relation $[X^1, P^+] = 0$ the operator $R=P^1/(P^0+P^z) $ is in the postulated algebra, 
\begin{equation}
\label{einsdurchpplus}
-\ir [X^1, P^-] = \frac{P^1}{P^0 + P^z}\ .
\end{equation}

In $D=3$ and $D=4$ this operator $R$ diverges on each smooth massless state $\Psi$ which does not 
vanish at a point $\hat p \in  \mathcal A_-$ on the negative $z$-axis
such that in some neighbourhood  $\mathcal U_{\hat p}$ of this point 
\begin{equation}
\label{lower2}
\text{for all } p\in \mathcal U_{\hat p}:\ |\Psi(p)|^2 > c > 0\ .
\end{equation}
If $\Psi\ne 0$ does not satisfy this condition  then there is a rotation $\rho$ such that
$U_\rho \Psi$ does, because $|U_\rho \Psi|(\hat p) = |\Psi|(\rho^{-1}\hat p)$ and each $p=\rho^{-1}\hat  p$, where
$\Psi$ does not vanish, can be rotated
to a point $\hat p$ on the negative $z$-axis.

Restrict for definiteness $\mathcal U_{\hat p}$ to a cylinder with $0 < a < -p^z< b$
and $0 < r < \bar r$, where~$r$ denotes the axial distance to the $z$-axis.

Proof: to determine the contribution of $\mathcal U_{\hat p}$ to $|R \Psi|^2$ we remark
that in cylindrical coordinates in $D=3$ the operator $P^1$ multiplies  with $r$ or $-r$,  
in $D=4$ it multiplies $\Psi(r\cos \theta,r\sin \theta,p^z)$ with $ r\cos \theta$. 
In $\mathcal U_{\hat p}$ the function $1/(p^0 + p^z)$ is larger than $2 |p^z|/r^2$ 
(\ref{lowerbound}). 
So in $U_{\hat p}$ the function $|\bigl(p^1/(p^0 + p^z)\bigr)\Psi(p)|^2$ is
larger in $D=3$ than $4 c |p^z|^2 (r/r^2)^2$ and in $D=4$ larger than $4 c|p^z|^2 (r/r^2)^2 \cos^2 \theta$.

The integral of this function in $D=4$ over the angle $\theta$ is a positive function of 
$r$ and~$(p^z)^2$, 
as the integrand vanishes as a function of the angle only in a set of spherical measure zero.
Then the integral over $p^z$ from $a$ to $b$ leaves a positive constant (the factor $1/p^0$ 
of the Lorentz invariant volume element $\dt p$ is larger than $1/\sqrt{b^2+\bar r^2}$)
times a linearly divergent integral $\int_{0}^{\bar r} \dv r /r^2$ in $D=3$, 
where $\dv^2 p = \dv r \dv p^z$, or in $D=4$ a logarithmically divergent integral 
$\int_{0}^{\bar r}\dv r\, r/r^2$ because $\dv^3 p \sim  r \dv r \dv p^z \dv \theta $.\qed

Consider the generators $-\ir M_{iz}$ of 
rotations in the $P^i$-$P^z$-plane. They commute with $P^j$,
$[-\ir M_{iz}, P^j]= 0$ if $j\ne i$ and $j\ne z$ and satisfy
\begin{equation}
 [-\ir M_{iz}, P^i] = P^z\ (\text{no sum over }i)\ ,\  [-\ir M_{iz}, P^z] = - P^i\ .
\end{equation}
To ease our subsequent discussion note that by the chain rule (valid outside $\mathcal A_-$)
\begin{equation}
[-\ir M_{1z}, \ln (P^0 + P^z)] = \frac{P^1}{P^0 + P^z} 
\end{equation}
the rough operator $P^1/(P^0+P^z)$ is the commutator of $-\ir M_{1z}$ with 
$\ln(P^0 + P^z) $ which is less rough.
Commutation with $-\ir M_{iz}$ increases roughness! Consider the $n$-fold commutator
of $M_{1z}, M_{2z}, \dots, M_{nz}$ with $\ln (P^0 + P^z)$ applied to $\Psi$ ($n\le D-2$)
\begin{equation}
\Psi_n =(-\ir)^n [M_{nz},\dots [M_{2z}, [M_{1z},\ln(P^0 + P^z)]]\dots]\Psi
=(n-1)!\,\frac{P^1\dots P^n}{(P^0+P^z)^n}\Psi  
\end{equation}
By (\ref{lowerbound},\ref{lower2}) $|\Psi_n|^2(p)$ is in $\mathcal U_{\hat p}$ larger than
\begin{equation}
\label{psi2bisk}
|\Psi_n|^2(p) > c ((n-1)!)^2 \left (\frac{r^n (2 |p_z|)^n}{r^{2n}}\right )^2 |f|^2
\end{equation}
where $
f$ is a function of the coordinates of the sphere $S^{D-3}$ of directions of the transverse momentum
and vanishes on the sphere only in a set of spherical measure zero (the coordinate planes). So the integral of
$|\Psi_n|^2(p)$ on  $S^{D-3}$ gives a positive number. 
The factor $1/p^0$ of the Lorentz invariant measure is larger than $1/\sqrt{b^2 + \bar r^2}$, 
the integral of $-p_z$ from $a$ to $b$ over this lower bound yields just a positive number.
The contribution of $\mathcal U_{\hat p}$ to the norm squared of $\Psi_n$ is therefore bounded
from below by a positive number times an integral
\begin{equation} 
\lim_{\varepsilon \rightarrow 0} \int_\varepsilon^{\bar r} \dv r \, \frac{r^{D-3}}{r^{2n}}\ .
\end{equation}
It diverges linearly for $D = 2n + 1$ and logarithmically for $D= 2n+2$.
For each $D> 2$ there is an $n$, such that the $(n-1)$-fold commutator of the 
generators $M_{2z}, M_{3z}, \dots M_{nz}$,
with $R=P^1/(P^0 + P^z)$ or the $n$-fold commutator of $M_{1z}, M_{2z}, \dots M_{nz}$,
with $\ln(P^0+P^z)$  cannot be applied to massless states $\Psi$ which satisfy (\ref{lower2}).
But for each $\Psi\ne 0$ there is a rotation $\rho$ such that $U_\rho\Psi$ satisfies 
(\ref{lower2}) and is not in the domain of the $n$-fold commutator.

The commutation relation, postulated by the canonical quantization of the string in the 
lightcone gauge are inconsistent with the existence of a massless unitary representation 
of the Poincar\'e group.
It is irrelevant that the operators, which are postulated in the canonical quantization 
of the lightcone string, are shown in an algebraic calculation  to satisfy in \mbox{$D=26$}
a Lorentz Lie algebra \cite{goddard} and to contain massless states. 
The operator $R=P^1/(P^0 + P^z)$, which is postulated in the canonical quantization,
and the generators of rotations have no common, massless domain.

Equally unimportant and possibly misleading is the observation \cite{fronsdal, sexl} that 
the measure  of the set $ \mathcal A_-$ vanishes where $p^1/(p^0 + p^z)$ diverges.
Essential is that $|\Psi_n(p)|^2$ (\ref{psi2bisk})
becomes so large in $\mathcal U_{\hat p}$  that in $D=2n+2$ or $D=2n + 1$ 
this function is not in the Hilbert space of square integrable states.

Though we have not worked out the details, analogous arguments should also exclude the excitation operators 
$\alpha_{-l}$ of section 3. They map massless states to massive states but cannot do so smoothly. 
They multiply with a determinantal factor $R$ (\ref{phil},\ref{psip}) which diverges at least 
in a point $\hat p$ of one of the involved mass shells. 
The Poincar\'e algebra always contains $(D-1-d)$ independent generators $-\ir M_{\omega}$ which act
like a derivative along some line through the $d$-dimensional manifold of points $\hat p$,
where $R$ becomes singular. Applied to a state, which does 
not vanish in $\mathcal U_{\hat p}$, we expect these derivatives to increase the order of 
the divergency until the $n$\textsuperscript{th} commutator applied to $\Psi$ gives a function
$\Psi_n$ which is not square integrable on $\mathcal U_{\hat p}$. The virtue of this arguments is, 
that it does not need $R$ to be in an algebra but considers only objects linear in $R$ as 
seems appropriate. After all, such a singularity $R$ occurs in the action of $\alpha_{-l}$ 
and $\alpha_l$ , $l\ge 2$, only once when they make the transition between the massive and 
massless shells.

\section{No Position Operator for Massless Particles}

Similar arguments exclude a position operator $\vec X$ 
which generates the group of translations of spatial momentum of massless particles
\begin{equation}
\label{transmom}
\bigl(\e^{\ir \vec b \cdot \vec X}\Psi\bigr)(\vec p)=\Psi(\vec p-\vec b)\ , \vec b \in \mathbb R^{D-1}\ .
\end{equation}
which is unitary relative to the measure $\dv^{D-1}\!p$.\footnote{In this section 
we absorb a factor $1/\sqrt{p^0}$ in the wavefunction and deal
with the translation invariant measure $\dv^{D-1}\!p$.} 
It enlarges the algebra of the Poincar\'e generators by Heisenberg partners $X^j$ of the spatial momenta,
\begin{equation}
\ [P^i, P^j]=0=[X^i, X^j]\ ,\ [P^i, X^j] = -\ir\, \delta^{ij}\ ,  \ i,j\in\set{1,\dots D-1}\ . 
\end{equation}

The Hamiltonian $H=P^0 = \sqrt{\vec P^2}$ is rough with respect to the translation group.
Consider a state $\Psi$ which in a ball
$\mathcal U_0=\set{p:\vec p^2 < \bar r ^2}$ around $0$ does not vanish
\begin{equation}
\label{lower3}
\text{ for all } p \in\mathcal U_0:\ |\Psi(p)|^2 > c > 0 
\end{equation}
and apply to it the $n+1$-fold commutator, $n < (D-2)$,
\begin{equation}
\Psi_n(p) = (-\ir)^{n+1}[X^{n+1}, \dots ,[X^{2},[X^{1}, \sqrt{\vec P^2}]]\dots ]\Psi(p)= 
c_n \frac{p^1p^2\dots p^{n+1}}{\sqrt{\vec p^2}^{2n+1}}\Psi(p)
\end{equation}
where $c_n$ is a non-vanishing combinatorial factor.
Its modulus squared is the product of a non-negative function $f$ of the spherical angles,
which vanishes on the sphere only in a set of vanishing spherical measure. So integrated over 
the sphere one is left as $\mathcal U_0$'s contribution to the norm squared of $\Psi_n$ with
a positive constant times an integral
\begin{equation} 
\lim_{\varepsilon \rightarrow 0} \int_\varepsilon^{\bar r} \dv r \, \frac{r^{D-2}}{r^{2n}}\ .
\end{equation}
To confirm this elementarily: each commutator with $X^i$
decreases for dimensional reasons the degree in $P$ by one. We started with $r=\sqrt{\vec p^2}$ 
of degree $1$ and obtain after \mbox{$n+1$} commutators degree $-n$, its modulus squared has 
degree $-2n$. In spherical coordinates $\dv^{D-1}\!p$ is 
of the form $r^{D-2} \dv r\dv \Omega_{D-2}$.

Analogous to the case of the rough operator $P^1/(P^0+P^z)$ we conclude, that the relativistic
massless Hamiltonian $H=\sqrt{\vec p^2}$ is incompatible with a Heisenberg algebra of 
$(D-1)$ spatial self-adjoint Heisenberg pairs.

The transverse position operators for massless particles exclude also the light cone string
in $D\ge 5$. It contains besides other operators $(D-2)$ spatial, transversal Heisenberg pairs
acting on massless states. In $D=5$ the wavefunction 
\begin{equation}
[-\ir X^3,[-\ir X^2,[-\ir X^1,\sqrt{\vec P^2}]]]\Psi(p) = 3 \frac{p^1 p^2p^3}{\sqrt{\vec p^2}^5}\Psi(p)
\end{equation}
has $r$-degree $-2$, if $\Psi(p)$ satisfies (\ref{lower3}). Its modulus squared behaves like
$1/r^4$. Integrated on $\mathcal U_0$ in sperical coordinates with 
$\dv^{4} p \sim r^3 \dv r \dv \Omega_3$ the $r$-integral diverges at $r=0$ logarithmically.
For each $\Psi\ne 0$ there is an $\e^{\ir \vec b \cdot \vec X}\Psi$, 
$\vec b\in \mathbb R^{D-1}$, which satisfies (\ref{lower3}).

All attempts \cite{hawton, newton, pryce, wightman} to construct position operators $\vec X$
for massless particles failed.

Different from massive particles the momentum spectrum  (not the Lorentz orbit) of  massless 
particles contains a Lorentz fixed point, $p = 0$. There the function $p^0=\sqrt{\vec p^2}$ of $\mathbb R^{D-1}$  is only continuous but not smooth. 
This single, distinguished 
point is sufficient to spoil the translation invariance of spatial momentum.
It prevents spatial translations $\e^{\ir \vec b\cdot \vec X}$ to enlarge on massless states 
the unitary action of the Poincar\'e group.
Repeated commutators of its generators~$\vec X$ with the Hamiltonian $P^0=\sqrt{\vec P^2}$
diverge on states $\Psi$, which satisfy (\ref{lower3}). 

\longpage

The proposal, to use the Fourier transformed (with 
respect to the spatial momentum) momentum wave function as position wave function,
does not work because $\Psi$ is a section (\ref{suednordwell}). The Fourier transformation of 
the local section $\Psi_N$ is not locally related to the one of $\Psi_S$.

That there is no position operator for massless particles disappoints expectations, 
because we see the world and reconstruct the position of all objects by light
which we receive as flow of massless quanta. But we do not see a distant photon. 
Rather we see massive objects by the 
currents of photons which they emit or scatter and which are annihilated in our retina.

Even if there is no position operator of massless particles we find the visual perception 
of the world sufficiently explained by the fact, that the luminosity $L$ of intersecting beams
is roughly proportional to the spacetime overlap 
\begin{equation}
\label{lumin1}
\begin{gathered}
L =  \frac{\sqrt{(p_1\cdot p_2)^2 - m_1^2m_2^2}}{p_1^0 p_2^0} \int\!\!\dv\!^{\,4}\! x\, 
 |\tilde \Psi_1(t,\vec x)|^2\,|\tilde \Psi_2(t,\vec x)|^2
\end{gathered}
\end{equation}
of the colliding wave packets $\tilde \Psi_1(x)$ and $\tilde \Psi_2(x)$  \cite{dragon}. 

\section{Conclusions}

Our investigation does not depend on this or that method of quantization but studies the resulting quantum theory. 
We exploit the smoothness of Lie groups. Their generators constitute an  algebra, which leaves
invariant a common domain, the smooth, rapidly decreasing wavefunctions. 
Smoothness and rapid decrease are properties which in bracket notation $\ket{p,i}$ 
are usually disregarded as it indicates not the state $\Psi: (p,i) \mapsto \Psi^i(p)$ 
but only its arguments. 

Canonical quantization of the light cone string postulates the operator
$P^1/(P^0+P^z)$. On massless states it is not smooth but diverges near the negative $z$-axis.
Repeated commutators of this operator with the generators of rotations in the $P^i$-$P^z$ planes,
$i=2, \dots n< D-1$, are shown to diverge on states which do not vanish on the negative $z$-axis.
 
Consequently the domain of $P^1/(P^0+P^z)$ does not contain all smooth states and is incompatible 
with a unitary representation of rotations.

Therefore the algebraic confirmation that in $D=26$ canonically quantized generators of 
the classical light cone string satisfy the Lorentz algebra is meaningless. 
In no dimension is $P^1/(P^0+P^z)$ 
defined on a complete Poincar\'e multiplet  of massless states. 

A related argument excludes on each massless multiplet a position operator $X^i$, 
which generates translations of the spatial momentum $P^j$, $i,j\in \set{1,\dots, D-1}$. 
The Hamiltonian $P^0=\sqrt{\vec P^2}$ is continuous but not continuously differentiable at $p=0$.
This momentum is not in the orbit of $\underline p=(1,0\dots, 1)$ but only a boundary point of the 
momentum spectrum. But it is rough enough to exclude a cohabitation on massless multiplets
of the Poincar\'e group and the group of spatial translations of momentum.

The operators $\alpha_l$ 
which perform \emph{momentum local} transitions between massless and massive states cannot 
map the natural domains of the Lorentz generators (the Schwartz spaces of their mass shells)
to each other, as the topologies of the shells differ. Our analysis of rough multiplicative
operators suggests that such operators cannot exist in relativistic theories.

On the other hand, the Lorentz generators acting on massless wavefunctions with non-vanishing he\-li\-ci\-ty 
(\ref{masselosnord}) show that rough multiplicative operators can combine with 
differential operators to smooth operators, the covariant derivatives (\ref{covdiv}).

\end{document}